\documentclass[a4paper]{article}
\usepackage{todonotes}
\usepackage{INTERSPEECH2021}
\usepackage{color, colortbl}
\usepackage{subfig}
\usepackage{url}
\usepackage{cite}

\usepackage{tipa} 

\usepackage{multirow}

\definecolor{Gray}{gray}{0.95}
\title{Quantifying Bias in Automatic Speech Recognition}
\name{Siyuan Feng$^1$, Olya Kudina$^2$, Bence Mark Halpern$^{1,3,4}$  and Odette Scharenborg$^1$}
\address{
  $^1$Multimedia Computing Group,
  $^2$Ethics and Philosophy of Technology Section VTI Department, Delft University of Technology, Delft, the Netherlands\\
  $^3$Netherlands Cancer Institute, Amsterdam, the Netherlands\\
  $^4$ACLC, University of Amsterdam, Amsterdam, the Netherlands}
\email{\{S.Feng,O.Kudina,O.E.Scharenborg\}@tudelft.nl, B.M.Halpern@uva.nl}

\begin{document}

\maketitle
\begin{abstract}
Automatic speech recognition (ASR) systems promise to deliver objective interpretation of human speech. Practice and recent evidence  suggests that the state-of-the-art (SotA) ASRs struggle with the large variation in  speech due to e.g.,  gender, age, speech impairment, race, and accents. Many factors can cause the bias of an ASR system. Our overarching goal is   to uncover bias in ASR systems to work towards proactive bias mitigation in ASR. This paper is a first step towards this goal and systematically quantifies the bias of a Dutch SotA ASR system against gender, age, regional accents and non-native accents.
Word error rates are compared, and an in-depth phoneme-level error analysis is conducted to understand where bias is occurring. We primarily focus on bias due to articulation differences in the dataset. Based on our findings, we suggest bias mitigation strategies for ASR development.
\end{abstract}
\noindent\textbf{Index Terms}: Quantifying bias, automatic speech recognition (ASR), gender, age, accent

\section{Introduction}
\label{sec:intro}
Automatic speech recognition (ASR) is increasingly used, e.g. in emergency response centers, domestic voice assistants, search engines, etc. Because of the paramount relevance spoken language plays in our lives, it is critical that ASR systems are able to deal with the variability in the way people speak (e.g., due to speaker differences, demographics, different speaking styles, and differently abled users). ASR systems promise to deliver objective interpretation of human speech. 

State-of-the-art ASR systems are based on deep neural networks (DNNs). DNNs are often considered to be a harbour of objectivity because they follow a clear path against the set parameters applied to the provided dataset. Although studies on bias in ASR are only nascent, practice and recent evidence is however troubling, suggesting that the state-of-the-art ASRs do not recognise the speech of everyone equally well. This evidence ranges from anecdotal (e.g., the Google Home of author O.S. typically does not recognise the speech of her 8-year-old daughter) to research- and policy-oriented. For instance, ASR systems have been shown to struggle with speech variance due to gender, age, speech impairment, race, and accents. Several studies on different languages have found gender differences: although most studies report that female speech is recognised better than male speech (Arabic \cite{shariah2013effects}, English \cite{koenecke2020racial,adda2005speech,goldwater2010words}, and French \cite{adda2005speech}),  the reverse pattern is also found (French \cite{garnerin2019gender}, English \cite{tatman2017gender}), although no difference in the recognition of  male and female speech was found in a follow-up study of the latter study \cite{tatman2017effects} nor was a difference found in \cite{garnerin2019gender}.
\cite{shariah2013effects} found that speakers younger than 30 years of age were better recognised than those older than 30 years. Moreover, ASR for child speech is proven more challenging than that for adult speech, due to children’s shorter vocal tracts, slower and more variable speaking rate and inaccurate articulation \cite{qian2017bidirectional}. 
A speech impairment is known to cause many problems for standard ASR systems, e.g., for impairments related to dysarthria \cite{laureano2019study}, stroke survival, oral cancer \cite{halpern2020detecting} or cleft lip and palate \cite{schuster2006evaluation}. Additionally, recent studies demonstrate how voice assistants perpetuate a racial divide by misrecognising the speech of black speakers more often than of white speakers \cite{koenecke2020racial,tatman2017effects}. Finally, ASR systems are typically trained on speech from native speakers of a “standard” variant of that language, inadvertently discriminating not only the speech of non-native speakers with high error rates \cite{wu2020see,palanica2019you} but also that of speakers of regional or sociolinguistic variants of the language (English \cite{koenecke2020racial,tatman2017gender,tatman2017effects}, Arabic \cite{shariah2013effects}). 

There are many factors that can cause this bias. First, the composition of the training data plays an important role. Moreover, a speaker with a type of language usage that deviates from the training data transcripts can lead to a mismatch with the language model. Articulation differences (i.e., differences in sound realisations) due to differences in speaking style, (regional/non-native) accent, vocal tract differences (e.g., due to gender, age) can lead to a mismatch between the speaker and the trained acoustic models (AMs). Additionally, a slower or faster speaking rate will result in a mismatch with the AMs. Another source of a possible bias is that the transcriptions can be biased. Anecdotal evidence (from author B.M.H. on the Jasmin-CGN  corpus \cite{cucchiarini2006jasmin}, see also Section \ref{subsec:error_analysis}) suggests that production errors of children are corrected (“normalised” towards what should have been said) in a more lenient way than those of non-native adult speakers (transcriptions tend to be more verbatim, including restarts), which leads to an increase in out-of-vocabulary (OOV) words and consequently an underestimation of the recognition performance for the latter group. Importantly, bias also creeps in far before the datasets are collected and deployed, e.g., when framing the problem, preparing the data and collecting it. Caliskan et al. showed that language corpora actually contain human-like biases \cite{caliskan2017semantics}. Moreover, possibly, bias can be due to the specific architectures and algorithms used in ASR system development. 

Powered by these concerns and equipped with a broad understanding of bias, our overarching goal in this project is to uncover bias in a standard DNN-based ASR system to work towards proactive bias-mitigation in ASR systems. In this paper, we systematically investigate the recognition performance on speech from different groups of speakers in order to quantify the bias in a standard, state-of-the-art Dutch ASR system\footnote{Code: \url{https://github.com/syfengcuhk/jasmin.}}. In other words, we investigate how well the ASR system can deal with the diversity in speech. In deviance to the above described work that typically focused on one to three dimensions, here we will investigate possible bias against gender, age (children, and older adults), regional accents and non-native accents. We compare word error rates (WERs), but will also carry out an in-depth analysis of which sounds are particularly prone to misrecognition in order to understand where bias is occurring. In this work, we focus on bias in the dataset, with a particular focus on bias due to articulation differences. Based on our findings, we will suggest potential bias mitigation strategies.
\section{Experimental set-up} 

\label{sec:exp_setup}
\subsection{Corpora}
\label{subsec:corpora}
\subsubsection{Dutch Spoken Corpus (CGN)}
The CGN corpus \cite{oostdijk2000spoken}  is used to train the standard-purpose ASR system  in this study. CGN contains Dutch recordings spoken by   speakers (age range 18-65 years old) from all over the Netherlands (NL) and Flanders (FL, in Belgium). It covers speaking styles including but not limited to read, broadcast news (BN)  and conversational telephone speech (CTS). In this study, CGN data from only NL is used, and its training and test data partition follows that of \cite{leeuwen2009results}. The total amount of training material is 483 hours, spoken by 1185 female and 1678 male speakers. 
\label{subsubsec:cgn}
\subsubsection{Jasmin-CGN corpus}
\label{subsubsec:jasmin}
The Jasmin-CGN corpus \cite{cucchiarini2006jasmin}, which is an extension of the CGN corpus, is used to evaluate the standard-purpose ASR system trained in this study on the dimensions of gender, age, regional and non-native accent\footnote{The training data of both CGN and Jasmin-CGN are recorded under 
a wide variety of recording conditions, which are potentially non-overlapping between the two corpora which might lead to an additional ASR performance deterioration on Jasmin-CGN.  }. 
Particularly, we use the speech from the following groups:
\begin{itemize}
    \item DC: native children; age 7--11; 12h 21m of speech;
    \item DT: native teenagers; age 12--16; 12h 21m of speech;
    \item DOA: native older adults; age 65+; 9h 26m of speech.
\end{itemize}
These speakers come from four different regions in the Netherlands: W: West, T: Transitional, N: North, S: South. Moreover, we were interested in testing the standard ASR trained on NL Dutch on another variant of Dutch: Flemish Dutch. Here, we follow the same age division as for the Dutch speakers: 
\begin{itemize}
    \item FC: Flemish children; age 7--11; 6h 10m of speech;
    \item FT: Flemish teenagers; age 12--16; 6h 10mm of speech;
    \item FOA: Flemish older adults; age 65+; 5h 5m of speech.
\end{itemize}
Table \ref{tab:num_spk} shows the number of speakers broken down by gender (female, male) for each age group and each region. In this study, FL is treated as a “region” similar to W, T, N, and S. 
\begin{table}[!t]
\renewcommand\arraystretch{0.7}
\centering
\caption{Number of native speakers (female, male) in each age group (C(hildren), T(eenagers), O(lder) A(dults)) per region in NL (W, T, N, S; indicated with D- in the first column) and for FL (indicated with F- in the first column).}
\resizebox{ 0.75 \linewidth}{!}{%
\begin{tabular}{l|ccccc}      
\toprule
Age & \multicolumn{5}{c}{Region}\\
  & W & T & N & S & FL\\
\midrule
DC/FC & 0,0 & 15,14 & 11,11& 9,11 & 23,19\\
DT/FT & 9,11 & 2,2 & 10,10 & 10,9 & 22,21\\
DOA/FOA & 13,5 & 9,8 & 13,4 & 10,6 & 21,16\\
\bottomrule
\end{tabular}%
}
\label{tab:num_spk}
\end{table}

Finally, we have two groups of non-native  speakers from the Netherlands, children and adults, with a wide range of native languages, including Turkish and Moroccan Arabic:
\begin{itemize}
    \item  NNC: non-native children; age 7--16; 12h 21m of speech;
    \item NNA: non-native adults; age 18--60; 12h 21m of speech.
\end{itemize}

Table  \ref{tab:num_spk_nonnative} shows the number of non-native children and adults broken down by gender (female, male), also separately for Dutch proficiency level according to the Common European Framework (CEF; A1 the lowest) for the adults. 

The Jasmin-CGN corpus consists of read speech and human-machine interaction (HMI) speech, both of which are used in the experiments.
\begin{table}[!t]
\renewcommand\arraystretch{0.7}
\centering
\caption{Number of non-native speakers (female, male) in each age group and CEF level (NNA) in the Jasmin-CGN corpus.}
\resizebox{ 0.65 \linewidth}{!}{%
\begin{tabular}{l|c|cccc}      
\toprule
Age &   \multicolumn{5}{c}{CEF level}\\
  & Sum& A1 & A2 & B1 & B2\\
\midrule
NNC & 28,25 & \multicolumn{4}{c}{N/A} \\
NNA & 28,17 & 4,6 &18,7 &6,3 & 0,1 \\
\bottomrule
\end{tabular}%
}
\label{tab:num_spk_nonnative}
\end{table}
\subsection{State-of-the-art ASR system for Dutch}
We adopt a hybrid DNN-HMM architecture \cite{dahl2011context} for training an ASR system, using   Kaldi   \cite{povey2011kaldi}. We tested with different mainstream DNN AM structures such as TDNNF, TDNN-LSTM and TDNN-BLSTM on the CGN test sets (BN and CTS) and found TDNN-BLSTM to be the best, thus TDNN-BLSTM is used throughout our experiments. The TDNN-BLSTM model consists of three TDNN layers of dimension 1024, and 3 pairs of forward-backward LSTM layers of cell dimension 1024 on top. The model is trained with the lattice-free maximum mutual information (LF-MMI) criterion \cite{povey2016purely}. We applied data augmentation techniques including speed perturbation \cite{ko2015audio}, reverberation \cite{ko2017study} and noise \cite{snyder2015musan} to the CGN training material, increasing the total hours of training data nine-fold, in order to increase our AM’s robustness towards different recording conditions in the evaluation data. 
The input features to the AM are 40-dimension high-resolution MFCCs. The AM is trained for 4 epochs. Context-dependent phone alignments used to train the AM are obtained by forced-alignment using a GMM-HMM trained beforehand with the same training data as that for the TDNN-BLSTM. 
The language model (LM) in our ASR system is an RNNLM \cite{xu2018neural}. It consists of 3 TDNN layers interleaved with  2 LSTM layers. 
To apply the RNNLM, a tri-gram LM is used to generate N-best results. After that, the RNNLM rescores the N-best results to get the final recognition results. The RNNLM and the tri-gram LM are   trained using the training data transcriptions in CGN.

\subsection{Experiments and Evaluation}
\label{subsec:exp_setup_eval}
In our experiments, the potential bias due to gender, age, regional and non-native accents is estimated for read speech and HMI speech separately. This allows us to investigate whether the size of the potential bias is influenced by the speaking style of the person. Read speech is typically well-articulated, and in general, ASR systems tend to perform well on read speech. HMI speech is less well prepared than read speech and possibly allows for more speaker-dependent articulations and differences in word usage, which might be more problematic for   ASR systems, and consequently have an influence on the size of the bias. 

The potential bias is estimated in terms of differences in WER between the different speaker groups. Additionally, we carry out an in-depth analysis at the phoneme level to investigate whether certain phonemes are prone to misrecognitions in order to investigate in how far atypical pronunciations are a possible source for bias to occur. To that end, we use a phoneme error rate (PER) based technique. The PER is calculated as follows: First, the word-level ground-truth and hypothesised (by the ASR system) transcripts are converted to phoneme-level sequences 
using the Dutch lexicon in CGN.
Second, the ground-truth and hypothesised phoneme sequences are aligned using the Levenshtein distance, after which the PER is calculated\footnote{Source code of the  analysis method can be found at: \url{https://github.com/karkirowle/relative_phoneme_analysis}.}.
\section{Results}
\label{sec:results}
\subsection{Baseline results}
\label{subsec:baseline}
Since there are no standard read speech and HMI test sets in CGN (which was used for training our ASR), the ASR system was first evaluated on the CGN standard BN and CTS test sets for reference. The ASR achieved 5.5\% WER on the BN set (female speech: 5.5\%; male speech: 5.4\%), and 20.8\% WER on the CTS set (female speech: 17.9\%; male speech: 23.2\%).

\subsection{Word recognition results}
\label{subsec:wer}
The WER averaged over all speakers was 36.2\% on read speech and 47.5\% on HMI speech. Table \ref{tab:wer_group} shows the WER per age group, for the female and male speech separately and averaged over both genders (column Avg), for read speech and HMI separately. The top rows report the results for the native Dutch speakers per age group; the bottom rows for the non-native speakers per age group. The WERs per gender, averaged over all age groups (row Avg), over the native (row AvgD) and non-native (row AvgN) Dutch speakers, respectively, are also shown.
\begin{table}[!t]
\renewcommand\arraystretch{0.7}
\centering
\caption{WERs on the read and HMI speech. “F/M” indicates female/male. “AvgD” indicates the average over all native Dutch speakers, AvgN over all non-native speakers and “Avg” indicates the average over all speakers.}
\resizebox{ 0.8 \linewidth}{!}{%
\begin{tabular}{l|cc>{\columncolor[rgb]{0.95,0.95,0.95}}c|cc>{\columncolor[rgb]{0.95,0.95,0.95}}c}      
\toprule
Group &   \multicolumn{3}{c|}{Read} & \multicolumn{3}{c}{HMI}\\
  & F& M & Avg & F & M & Avg\\
\midrule
DC & 34.8 & 35.7 &35.3 &43.5 &43.3 & 43.4  \\
DT &  16.5&20.1&18.4&34.4&36.2&35.3 \\
DOA &  22.3&27.9&24.2&37.8&42.5&39.5\\
\rowcolor{Gray}
AvgD & 24.4&28.1&26.1&38.4&41.7&39.8\\
\midrule
NNC  &54.3&55.9&55.1&60.9&62.1&61.6\\
NNA &57.3&56.1&56.9&61.2&61.5&61.3\\
\rowcolor{Gray}
AvgN  &55.8&56.0&55.9&61.1&61.7&61.4\\
\midrule
\rowcolor{Gray}
Avg &35.4&37.2&36.2&46.5&49.0&47.5\\
\bottomrule
\end{tabular}%
}
\label{tab:wer_group}
\end{table}
\begin{figure}[!t]
\centering
\subfloat[Read speech]{

    \includegraphics[width=\linewidth]{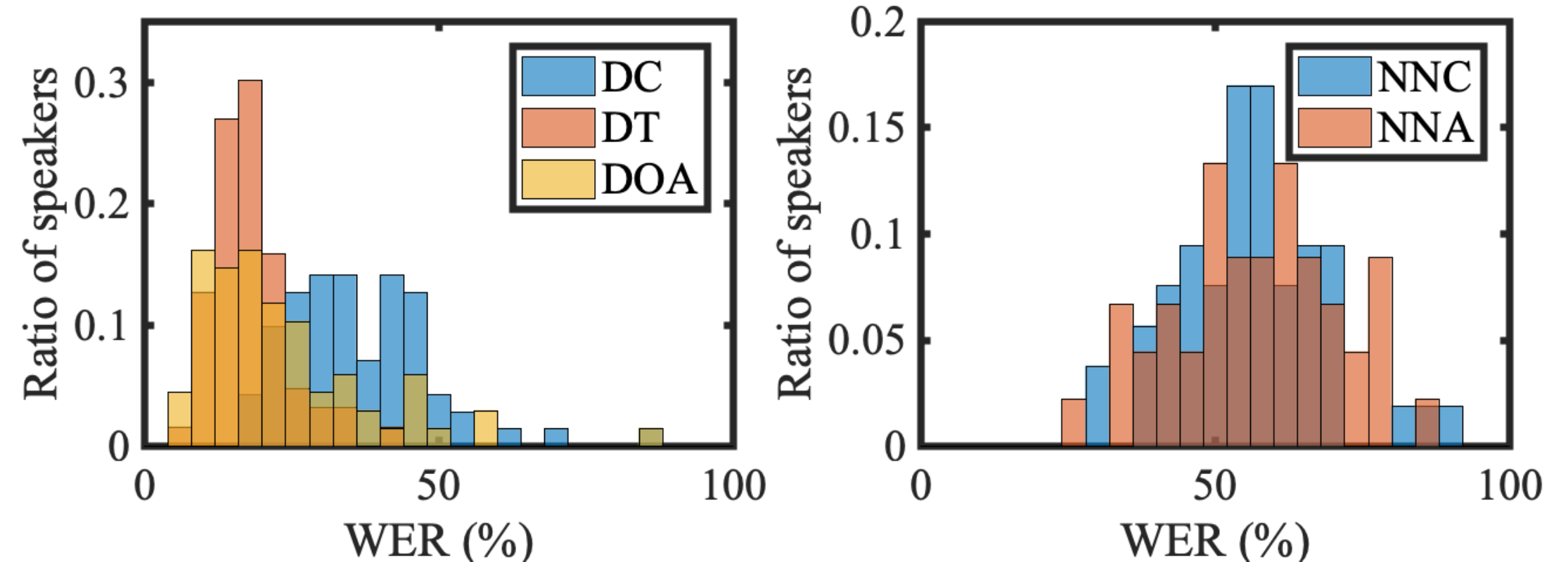}
    \label{hist_read}
   
}    \\
\subfloat[HMI speech]{

    \includegraphics[width=\linewidth]{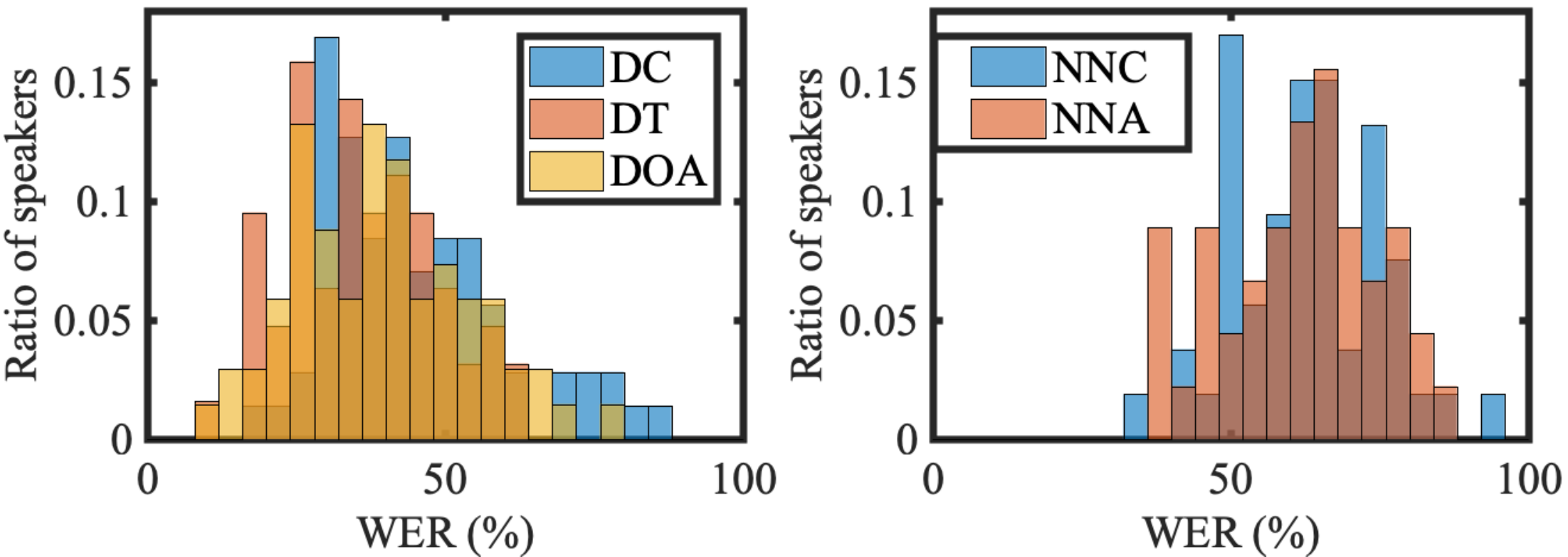}
    \label{hist_hmi}
   
}    
 \caption{Per-speaker WER histogram of read (a) and HMI (b) speech. Left: native speakers; right: non-native speakers; bin size: 4\%.}
  \label{fig:per_spk_wer}
\end{figure}

Table \ref{tab:wer_group} shows that, in general, female speech is better recognised than male speech. This is true for all native and non-native  groups and for both speech styles. The female-male WER difference is the largest in DOA and the smallest in DC, for both the read and HMI speech styles. 

Looking at the different age groups, Table \ref{tab:wer_group} shows that among the native speakers, DT achieves the best WER performances in read and HMI speech, followed by the DOA, while DC was the worst recognised. Among the non-native speakers, the performance differences between NNC and NNA do not differ much (absolute 1.8\% and 0.3\% in read and HMI speech, respectively). 
To gain a better understanding of the WER of the different age groups, Figure  \ref{fig:per_spk_wer} illustrates the per-speaker  WER histogram for read speech (a) and HMI speech (b) of these groups. Figure \ref{hist_read} shows for read speech, speaker-level WERs in DOA are more variable than in DT. Manual checking of some DOA speakers that had high per-speaker read speech WERs ($>$50\%) suggested that  their speech was not that well articulated (possibly due to their (old) age ($>$ 75)). Comparing to read speech, for HMI speech, there are less differences in histogram of groups DC, DT and DOA.
Figure \ref{fig:per_spk_wer} also shows the per-speaker WER histogram of the two non-native groups does not differ much, for both the read and HMI speech.

Comparing the native (D-) with the non-native (NN-) groups shows that speech of native speakers is recognised much better than that of non-native speakers of Dutch. The worst recognised native speech (DC; Dutch children)  has a read speech WER that is around 20\% absolute better than that of the best non-native age group (NNC; non-native children).

Table \ref{tab:wer_cef} provides a closer look at the WERs for the different Dutch proficiency levels (CEF) of the non-native adult speakers (NNA), separated by gender. 
\begin{table}[!t]
\renewcommand\arraystretch{0.7}
\centering
\caption{WERs of group NNA by CEF levels (A1 the lowest level). “F/M'' indicates female/male. The one B2-level speaker is omitted from NNA.}
\resizebox{ 0.75 \linewidth}{!}{%
\begin{tabular}{l|cc>{\columncolor[rgb]{0.95,0.95,0.95}}c|cc>{\columncolor[rgb]{0.95,0.95,0.95}}c}      
\toprule
CEF &   \multicolumn{3}{c|}{Read} & \multicolumn{3}{c}{HMI}\\
  & F& M & Avg & F & M & Avg\\
\midrule
A1 &58.4&56.4&57.2&65.8&62.7&63.4 \\
A2 & 58.3& 51.4& 56.5&61.6&56.6&60.3\\
B1 & 53.3&64.4&57.0&59.5&63.5&60.5\\
 
\bottomrule
\end{tabular}%
}
\label{tab:wer_cef}
\end{table}
Perhaps surprisingly, we do not see a reduction in WER with an increase in CEF level. 

Finally, Tables \ref{tab:wer_group} and \ref{tab:wer_cef} show that for each group, the WER performance of HMI speech is consistently worse than that of read speech. Overall, the absolute WER difference between read speech and HMI speech is around 13.7\% for native speakers, and is around 5.5\% for non-native speakers of Dutch.

Table \ref{tab:wer_region} shows the WERs with regards to regional accents of the four large regions in the Netherlands (W, T, N and S) and Flanders (FL) per age group. The average WER results of read speech (\ref{tab:wer_region_read}) and HMI speech (\ref{tab:wer_region_hmi}) in each age group are shown in the gray rows. and the WER results broken down by gender (female, male) are shown in the white rows. Table \ref{tab:wer_region} shows that speech spoken by people from Flanders (FL) achieved the worst WER performance in all age groups except for the older adults (DOA/FOA) (S the worst). This is for both read and HMI speech. For read speech, among the four regions in the Netherlands, no region was consistently recognised worse than others. For HMI speech, region S in general was the worst recognised. 

Looking at the Dutch age groups, Table \ref{tab:wer_region_read} shows that for the children and teenagers (DC and DT), the read speech differences in WER between the four regions vary much less (5\%@DC, 6\%@DT) than for the older Dutch speakers (DOA, 19\%). The same observation is made for HMI speech in Table \ref{tab:wer_region_hmi} (2\%@DC, 8\%@DT, 18\%@DOA). This suggests that older speakers in the Netherlands typically have stronger regional accents than children and teenagers. Specifically, in DOA, region S has the highest WER, and within this group, male speech is worse recognised than female speech by an absolute WER difference of 8\% for read speech and 3.3\% for HMI speech.


\begin{table}[!t]
\renewcommand\arraystretch{0.7}
\centering
\caption{WERs of read   (a) and HMI (b) speech  of the four Dutch regions and Flanders per age group. The average WERs are shown in the gray rows, and the WERs broken down by gender (female, male) are shown in the white rows.}

 \subfloat[Read speech]%
       {\resizebox{ 0.9 \linewidth}{!}{%
\begin{tabular}{l|cccccccccc}      
\toprule
Group & W&T&N&S&FL\\
 \midrule
 \multirow{ 2}{*}{DC/FC }  & \cellcolor[rgb]{0.95,0.95,0.95} N/A &\cellcolor[rgb]{0.95,0.95,0.95} 33.2 & \cellcolor[rgb]{0.95,0.95,0.95}38.2 & \cellcolor[rgb]{0.95,0.95,0.95}34.8 & \cellcolor[rgb]{0.95,0.95,0.95}52.1 \\
  & N/A & 30.7,35,6 & 35.9,40.4 & 41.0,30.1 & 48.4,56,4\\
  \midrule
 \multirow{ 2}{*}{DT/FT}  & \cellcolor[rgb]{0.95,0.95,0.95} 19.3 &\cellcolor[rgb]{0.95,0.95,0.95} 22.9 & \cellcolor[rgb]{0.95,0.95,0.95}16.8 & \cellcolor[rgb]{0.95,0.95,0.95}17.9 & \cellcolor[rgb]{0.95,0.95,0.95}40.5\\
 &16.9,21.2&18.9,26.3&15.9,17.7&16.2,19.6&38.2,43.2\\

 \midrule
  \multirow{ 2}{*}{DOA/FOA}  &\cellcolor[rgb]{0.95,0.95,0.95} 20.8  &\cellcolor[rgb]{0.95,0.95,0.95} 23.7 & \cellcolor[rgb]{0.95,0.95,0.95}17.0 & \cellcolor[rgb]{0.95,0.95,0.95}35.7 & \cellcolor[rgb]{0.95,0.95,0.95}29.9 \\
 & 18.0,27.7&24.3,22.9& 16.7,18.0& 32.7,40.7& 28.7,31.5\\
\bottomrule
\end{tabular}%
}
\label{tab:wer_region_read}

}

\subfloat[HMI speech]%
       {\resizebox{ 0.9 \linewidth}{!}{%
\begin{tabular}{l|cccccccccc}      
\toprule
Group & W&T&N&S&FL\\
 \midrule
 \multirow{ 2}{*}{DC/FC }  &\cellcolor[rgb]{0.95,0.95,0.95} N/A &\cellcolor[rgb]{0.95,0.95,0.95}  43.7 &  \cellcolor[rgb]{0.95,0.95,0.95} 42.2 & \cellcolor[rgb]{0.95,0.95,0.95} 43.5 & \cellcolor[rgb]{0.95,0.95,0.95} 65.0\\
  &  N/A & 42.2,45.2 &40.7,44.4 & 48.2,39.9 &  66.4,63.5\\
  \midrule
 \multirow{ 2}{*}{DT/FT} & \cellcolor[rgb]{0.95,0.95,0.95}  32.8 &\cellcolor[rgb]{0.95,0.95,0.95}  33.0 & \cellcolor[rgb]{0.95,0.95,0.95} 32.7 & \cellcolor[rgb]{0.95,0.95,0.95} 41.1 & \cellcolor[rgb]{0.95,0.95,0.95} 50.1\\
 &29.2,36.7 & 33.0,32.2 & 32.1,33.4& 42.6,39.3 & 48.3,53.1\\

 \midrule
  \multirow{ 2}{*}{DOA/FOA}  & \cellcolor[rgb]{0.95,0.95,0.95} 37.6  &\cellcolor[rgb]{0.95,0.95,0.95} 36.7 & \cellcolor[rgb]{0.95,0.95,0.95} 31.8 & \cellcolor[rgb]{0.95,0.95,0.95} 49.7 & \cellcolor[rgb]{0.95,0.95,0.95} 48.5\\
 & 32.4,44.6&37.6,35.9&29.5,42.1 & 48.6,51.9 & 47.8,48.8\\
\bottomrule
\end{tabular}%
}
\label{tab:wer_region_hmi}
}
\label{tab:wer_region}

\end{table}
\subsection{Error analyses}
We analyse the sources of the recognition errors by the ASR    through a   systematic analysis of the phoneme errors, and  a qualitative analysis of the dataset, supplemented by post-hoc quantitative results when appropriate.  We first report the general findings and then our four main variables are  assessed:  non-native accents, age groups, regional accents, and gender.

In general, the PER of /\textipa{\textltailn}, /\textipa{S}/ and /\textipa{Z}/ seems to be consistently high for all age groups, however, these phonemes occur rarely ($<$50) in the data of most groups. 
To account for this, we only report \textbf{top-5} phonemes where there are at least 50 occurrences and do not report these three phonemes.

First, regarding non-native v.s. native accents, we find that the top-5 misrecognised phonemes in group NNA for A1 are /\textipa{\oe y}/, /\textipa{Y}/, /\textipa{y}/, /\o:/, /h/; For A2: /\textipa{\oe y}/, /\textipa{Y}/, /\textipa{y}/, /\textipa{j}/; For B1: /\textipa{\o ey}/, /\o:/, /\textipa{Y}/, /\textipa{h}/, /\textipa{j}/. For native (older) speakers (DOA), these phonemes are /\textipa{h}/, /\textipa{x}/, /\textipa{E}/, /\textipa{@}/. The results show that different sounds are difficult to recognise for the ASR for the non-native and native speakers. Particularly vowels that are known to be difficult to acquire for many second-language learners of Dutch are badly recognised, i.e., /\textipa{\oe y}/, /\textipa{Y}/, /\textipa{y}/, and /\o:/.


Looking at the different age groups of the native Dutch speakers, we find the top-5 misrecognised phonemes for DC are: /\textipa{Y}/, /\textipa{h}/, /\textipa{@}/, /\textipa{j}/; 
For DT:  /\textipa{Y}/, /\o:/, /\textipa{h}/, /\textipa{@}/;
For DOA: /\textipa{h}/, /\textipa{x}/, /\textipa{E}/, /\textipa{@}/. 
While /\textipa{S}/ was difficult to recognise in most condition for the ASR system, we've noticed that in the case of native children (DC, age 7-11), this issue is exacerbated, and there are  sibilant pronunciations that confuse the ASR system. This is confirmed by certain substitution errors (sullen vs zullen, zal vs saul) in the decoded sentences. 



For the regional variants of Dutch, the breakdown of the PER by regions W, T, N, S and FL is as follow: 
W: /\textipa{h}/, /@/, /\textipa{Y}/, /\textipa{E}/; 
T: /\textipa{h}/, /\textipa{z}/, /\textipa{@}/, /\textipa{Y}/; 
N: /\textipa{h}/, /\textipa{Y}/, /\textipa{O}/, /\textipa{z}/; 
S:   /\textipa{h}/, /\textipa{x}/, /\textipa{E}/, /\textipa{@}/;
FL: /\textipa{y}, /\textipa{\oe y}/, /\textipa{Au}/.
Two patterns can be observed in this data: 1) the vowels that were problematic for in non-native speech also appear as the most often recognised phonemes for several of the Dutch and Flemish regions. This suggests that also within the native speakers, these vowels have a large variation in their production. 2) The DOA of region S were shown to have the highest WER (see Table \ref{tab:wer_region}). This can be explained by the high misrecognition rate of /\textipa{x}/, which is known to be produced differently in the southern provinces  of NL compared to “standard” Dutch. 


Finally, regarding gender, the top-5 misrecognised phonemes for male speakers are:  /\o:/, /\textipa{Au}/, while for female speakers, these are:  /\textipa{\o ey}/, /\textipa{Y}/, /\textipa{h}/, /\textipa{\o:}/. There is no clear pattern in terms of misrecognised phonemes although we do observe that overall, male speech seems to achieve consistently higher PERs than female speech irrespective of the phoneme identity.


\label{subsec:error_analysis}
\section{General discussion and conclusion}
In this paper, we have shown that an ASR system can perpetuate the existing bias in society. We have quantified bias in a state-of-the-art standard Dutch ASR system with regards to diversity in gender, age, regional accents and non-native accents.

We found that female speech was better recognised than male speech. This result adds to a growing set of findings that male and female speech are not recognised equally well \cite{koenecke2020racial,tatman2017gender,shariah2013effects,adda2005speech,goldwater2010words}. Teenagers’ speech is the best recognised, followed by senior people’s (over 65 y/o) and children’s speech is the worst. 
The problems of the ASR with recognising children’s speech are not surprising: the large difference in children’s speech and adults’ speech \cite{qian2017bidirectional}  leads to a large mismatch of the children’s speech with the AM.
The worse recognition of the older adults' speech, especially those over 75 y/o,  is due to a less well articulation. 
Possibly, the speech of the teenagers resembles the speech of the adult speakers in CGN the most.

Speech of native Dutch speakers is much better recognised than that of non-native speakers, irrespective of age. This is in line with the qualitative findings reported in \cite{wu2020see,palanica2019you}, and is not surprising: Non-native speakers typically have an accent, meaning that the match with the AM is worse than that of native speakers. Interestingly, for the non-native speakers, no correlation was found between Dutch proficiency level and the ASR performance. A reason might be that at the A1, A2 and B1 levels the focus is primarily on vocabulary and grammar rather than pronunciation, so that the pronunciation only differs little between the different proficiency levels. Another reason might be that the proficiency level in general is not a good proxy for the strength of the accent.

For native Dutch speakers, the speech  from Flanders (FL) obtained the worst ASR performance and worse than   all the regions in the Netherlands. The much higher WER results for the FL speech is well explained by the large accent difference between Dutch spoken in the Netherlands (used to train the ASR system) and that spoken in Flanders.
For regions within the Netherlands, we found that regional accents seem to be stronger for older people than children and teenagers as their speech was recognised worse than that of children and teenagers. 

Finally, we found HMI speech  to be consistently worse recognised than read speech. This confirms that the size of the bias is influenced by the speaking style of the person. Indeed, potentially HMI speech, which is less well prepared than read speech allows for more speaker-dependent articulations and differences in word usage, which cause problems for the ASR system.


The above results show that the composition of the training data plays an important role in the performance difference of an ASR system for a diverse range of speech.

In this paper, we have focused on bias that can be quantified. 
However, owing to the foundational nature of bias,
it is impossible to remove bias that creeps into datasets \cite{kudina2021co}. This becomes a priority in responsible ASR system development: framing the problem, developing the developer team composition and the implementation process from a point of anticipating, proactively spotting, and developing mitigation strategies for prejudice.
A direct   bias mitigation strategy concerns diversifying and aiming for a balanced representation of all types of speakers in the dataset \cite{koenecke2020racial,caliskan2017semantics}. 
An indirect bias mitigation strategy deals with diverse team composition: the variety in age, regions, gender, etc. provides additional lenses of spotting   potential bias in design. Together, they can help 
ensure a more inclusive developmental environment for ASR. 
In conclusion, the general impression of ASR systems being biased is now supported by more data.
 
\section{Acknowledgements}

B.M.H. is funded through the EU’s H2020 research and innovation programme under MSC grant agreement No 766287. 

\bibliographystyle{IEEEtran}

\bibliography{mybib}


\end{document}